\documentclass[runningheads]{llncs}
\usepackage{epsfig}
\usepackage{amsmath}
\usepackage{amssymb}
\usepackage{multirow}
\usepackage{booktabs}
\usepackage{authblk}
\usepackage{blindtext}
\usepackage{xcolor}
\usepackage{makecell}
\usepackage{subcaption}

\newcommand{\ie}{\textit{i}.\textit{e}.}
\newcommand{\eg}{\textit{e}.\textit{g}.}

\usepackage{graphicx}

\makeatletter
\newcommand{\printfnsymbol}[1]{%
  \textsuperscript{\@fnsymbol{#1}}%
}
\makeatother

\begin{document}
\title{Efficient and Phase-aware Video Super-resolution for Cardiac MRI}
\author{Jhih-Yuan Lin \thanks{equal contribution} \and Yu-Cheng Chang \printfnsymbol{1} \and Winston H. Hsu}
\authorrunning{J.Y. Lin et al.}
\institute{National Taiwan University, Taipei, Taiwan}
\maketitle

\begin{abstract}
Cardiac Magnetic Resonance Imaging (CMR) is widely used since it can illustrate the structure and function of the heart in a non-invasive and painless way.
However, it is time-consuming and high-cost to acquire high-quality scans due to the hardware limitation.
To this end, we propose a novel end-to-end trainable network to solve CMR video super-resolution problem without the hardware upgrade and the scanning protocol modifications.
We incorporate the cardiac knowledge into our model to assist in utilizing the temporal information.
Specifically, we formulate the cardiac knowledge as the periodic function, which is tailored to meet the cyclic characteristic of CMR.
Besides, the proposed residual of residual learning scheme facilitates the network to learn the LR-HR mapping in a progressive refinement fashion.
This mechanism enables the network to have the adaptive capability by adjusting refinement iterations depending on the difficulty of the task.
Extensive experimental results on large-scale datasets demonstrate the superiority of the proposed method compared with numerous state-of-the-art methods.
\keywords{Cardiac MRI \and Video super-resolution}
\end{abstract}

\section{Introduction}
Magnetic Resonance Imaging (MRI) has been widely used to examine almost any part of the body since it can depict the structure inside the human non-invasively and produce high contrast images.
Notably, cardiac MRI (CMR) assessing cardiac structure and function plays a key role in evidence-based diagnostic and therapeutic pathways in cardiovascular disease~\cite{von2017representation}, including the assessment of myocardial ischemia, cardiomyopathies, myocarditis, congenital heart disease~\cite{von2015role}.
However, obtaining high-resolution CMR is time-consuming and high-cost as it is sensitive to the changes in the cardiac cycle length and respiratory position~\cite{salerno2017recent}, which is rarely clinically applicable.
To address this issue, the single image super-resolution (SISR) technique, which aims at reconstructing a high-resolution (HR) image from low-resolution (LR) one, holds a great promise that does not need to change the hardware or scanning protocol.
Most of the MRI SISR approaches~\cite{pham2017brain,chen2018efficient,shi2018super} are based on the deep learning-based methods~\cite{dong2015image,ledig2017photo}, which learn the LR-HR mapping with extensive LR-HR paired data.
On the other hand, several previous studies~\cite{jog2016self,zhao2018self} adapt the self-similarity based SISR algorithm~\cite{huang2015single}, which does not need external HR data for training.
However, straightforwardly employing the aforementioned methods is not appropriate for CMR video reconstruction since the relationship among the consecutive frames in CMR video is not well considered.
Therefore, we adopt the video super-resolution (VSR) technique, which can properly leverage the temporal information and has been applied in numerous works~\cite{sajjadi2018frame,jo2018deep,xue2019video,wang2019edvr,haris2019recurrent}, to perform CMR video reconstruction.

\begin{figure}[t]
    \centering
    \includegraphics[width=0.97\linewidth,trim={0cm 6cm 0cm 4cm},clip]{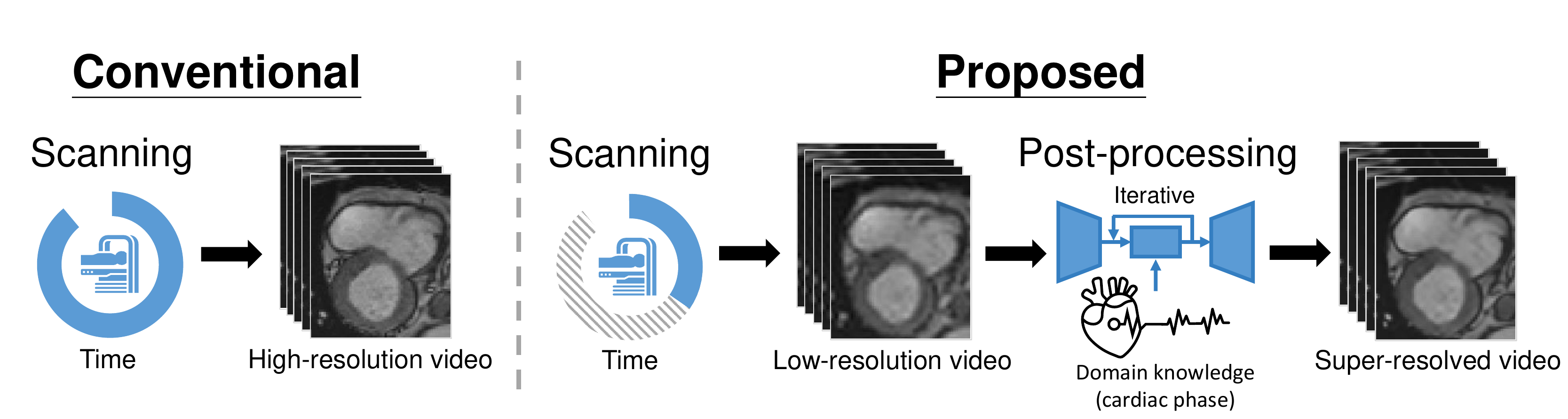}
    \caption{
        We present efficient post-processing to facilitate the acquisition of high-quality cardiac MRI (CMR) that is conventionally time-consuming, high-cost, and sensitive to the changes in the cardiac cycle length and respiratory position~\cite{salerno2017recent}.
        Specifically, we utilize the domain knowledge and iteratively enhance low-resolution CMR by a neural network, which can reduce the scan time and cost without changing the hardware or scanning protocol.
    }
    \label{fig:figure1}
\end{figure}

In this work, we propose an end-to-end trainable network to address CMR VSR problem.
To well consider the temporal information, we choose ConvLSTM~\cite{xingjian2015convolutional}, which has been proven effective~\cite{finn2016unsupervised,huang2015bidirectional}, as our backbone.
Moreover, we introduce the domain knowledge (\ie, cardiac phase), which has shown to be important for the measurement of the stroke volume~\cite{lalande2004left} and disease diagnosis~\cite{xu2017volume}, to provide the direct guidance about the temporal relationship in a cardiac cycle.
Combined with the proposed \emph{phase fusion module}, the model can better utilize the temporal information.
Last but not the least, we devise the \emph{residual of residual learning} inspired by the iterative error feedback mechanism~\cite{mnih2013playing,carreira2016human} to guide the model iteratively recover the lost details.
Different from other purely feed-forward approaches~\cite{lim2017enhanced,jo2018deep,wang2019edvr,xue2019video,sajjadi2018frame}, our iterative learning strategy can make the model easier in representing the LR-HR mapping with fewer parameters.
We evaluate our model and multiple state-of-the-art baselines on two synthetic datasets established by mimicking the acquisition of MRI~\cite{chen2018brain,zhao2018self} from two publicly datasets~\cite{bernard2018deep,datasciencebowl2015}.
It is worth noting that one of them is totally for external evaluation.
To properly assess the model performance, we introduce the cardiac metrics based on PSNR and SSIM.
The experimental results turn out that the proposed network can stand out from existing methods even on the large-scale external dataset, which indicates our model has the generalization ability.
To our best knowledge, this work is the pioneer to address the CMR VSR problem and provide a benchmark to facilitate the development in this domain.

\begin{figure*}[t]
    \centering
    \includegraphics[width=\linewidth,trim={0 3.8cm 0 4.2cm},clip]{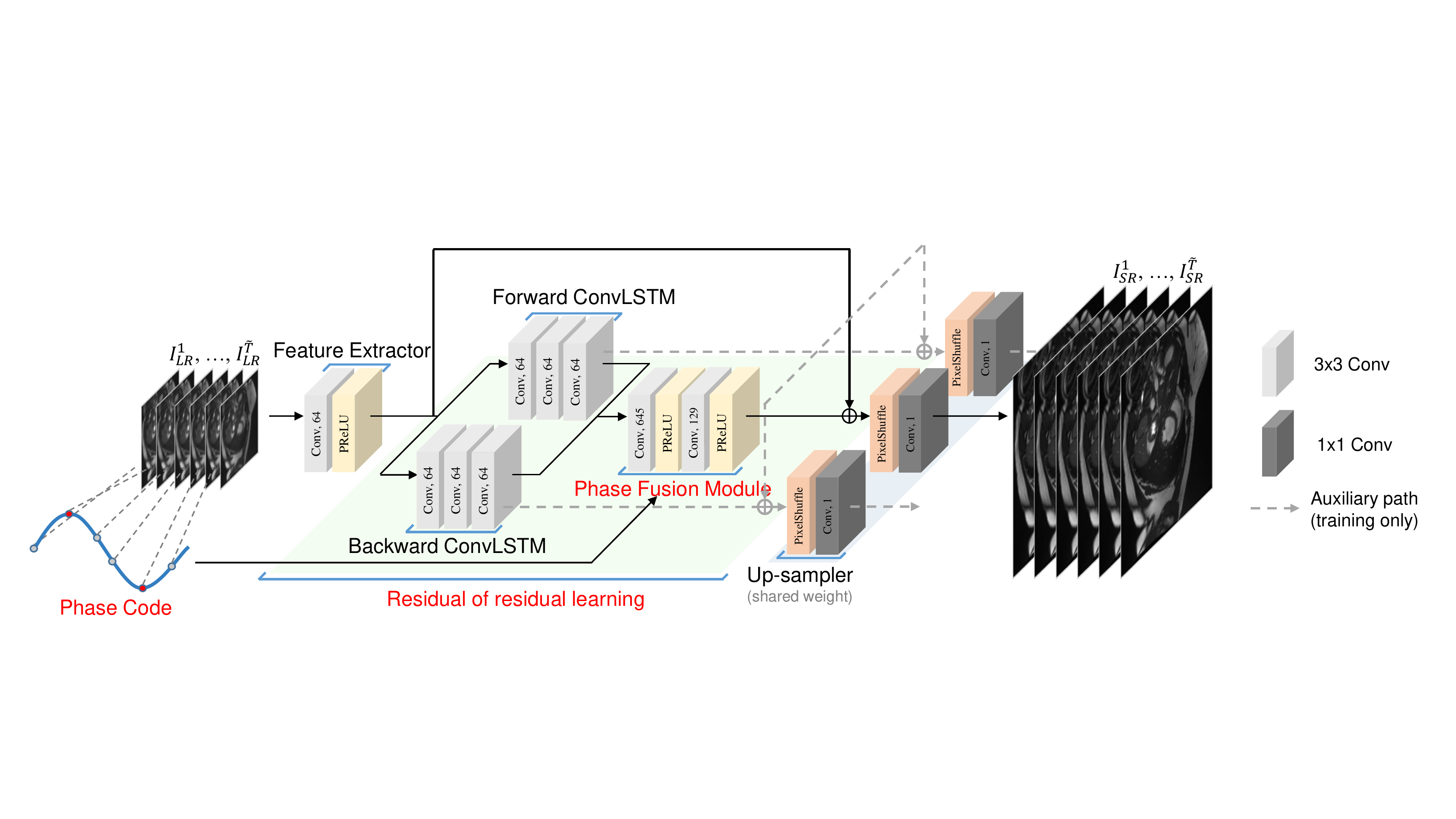}
    \caption{
        \textbf{Model overview.}
        The bidirectional ConvLSTM~\cite{xingjian2015convolutional} utilizes the temporal information from forward and backward directions.
        The \emph{phase fusion module} exploits the informative phase code to leverage the bidirectional features.
        With the \emph{residual of residual learning}, the network recovers the results in a coarse-to-fine fashion.
        Auxiliary paths are adopted for stabilizing the training procedure.
    }
    \label{fig:network}
\end{figure*}

\section{Proposed approach}
Let $I_{LR}^t$ $\in \mathbb{R}^{H \times W}$ denote the $t$-th LR frame obtained by down-sampling the original HR frame $I_{HR}^t$ $\in \mathbb{R}^{rH \times rW}$ with the scale factor $r$.
Given a sequence of LR frames denoted as \{$I_{LR}^t$\}, the proposed end-to-end trainable model aims to estimate the corresponding high-quality results \{$I_{SR}^t$\} that approximate the ground truth frames \{$I_{HR}^t$\}.
Besides, $\oplus$ refers to the element-wise addition.

\subsection{Overall architecture}
\label{sec:overall architecture}
Our proposed network is illustrated in Fig~\ref{fig:network}.
It consists of a feature extractor, a bidirectional ConvLSTM~\cite{xingjian2015convolutional}, a phase fusion module, and an up-sampler.
The feature extractor ($FE$) first exploits the frame $I_{LR}^t$ to obtain the low-frequency feature $L^t$.
Subsequently, the bidirectional ConvLSTM~\cite{xingjian2015convolutional} comprising a forward ConvLSTM ($ConvLSTM_F$) and a backward ConvLSTM ($ConvLSTM_B$) makes use of the low-frequency feature $L^t$ to generate the high-frequency features $H^t_F, H^t_B$.
With the help of its memory mechanism, the bidirectional ConvLSTM can fully utilize the temporal relationship among consecutive frames in both directions.
In addition, we can update the memory cells in the bidirectional ConvLSTM in advance instead of starting with the empty states due to the cyclic characteristic of the cardiac videos.
This can be done by feeding $n$ consequent updated frames before and after the input sequence \{$I^t_{LR}$\} to the network.
Furthermore, to completely integrate the bidirectional features, the designed phase fusion module ($PF$) applies the cardiac knowledge of the $2N+1$ successive frames from $t-N$ to $t+N$ in the form of the phase code $P^{[t-N:t+N]}$, which can be formulated as $H_P^t = PF(H^{[t-N:t+N]}_F, H^{[t-N:t+N]}_B, P^{[t-N:t+N]})$, where $H_P^t$ represents the fused high-frequency feature.
After that, the fused high-frequency feature $H_P^t$ combined with the low-frequency feature $L^t$ through the global skip connection is up-scaled by the up-sampler ($Up$) into the super-resolved image $I^t_{SR} = Up(H_P^t \oplus L^t)$.
We further define the sub-network ($Net_{sub}$) as the combination of $ConvLSTM_F, ConvLSTM_B$ and $PF$.
The purpose of $Net_{sub}$ is to recover the high-frequency residual $H_P^t = Net_{sub}(L^t)$.
Besides, we employ the deep supervision technique~\cite{lee2015deeply} to provide the additional gradient signal and stabilize the training process by adding two auxiliary paths, namely $I^t_{SR, F} = Up(H_F^t \oplus L^t)$ and $I^t_{SR, B} = Up(H_B^t \oplus L^t)$.
Finally, we propose the residual of residual learning that progressively restores the residual that has yet to be recovered in each refinement stage $\omega$.
To simplify the notation, $\omega$ is omitted when it equals to $0$, \eg, $L^t_F$ means the low-frequency feature of the $t$-th frame at the $0$-th stage $L^{t, 0}_F$.

\begin{figure}[t]
    \centering
    \begin{subfigure}{0.18\linewidth}
        \centering
        \begin{subfigure}{\linewidth}
            \begin{subfigure}{\linewidth}
                \includegraphics[width=\linewidth]{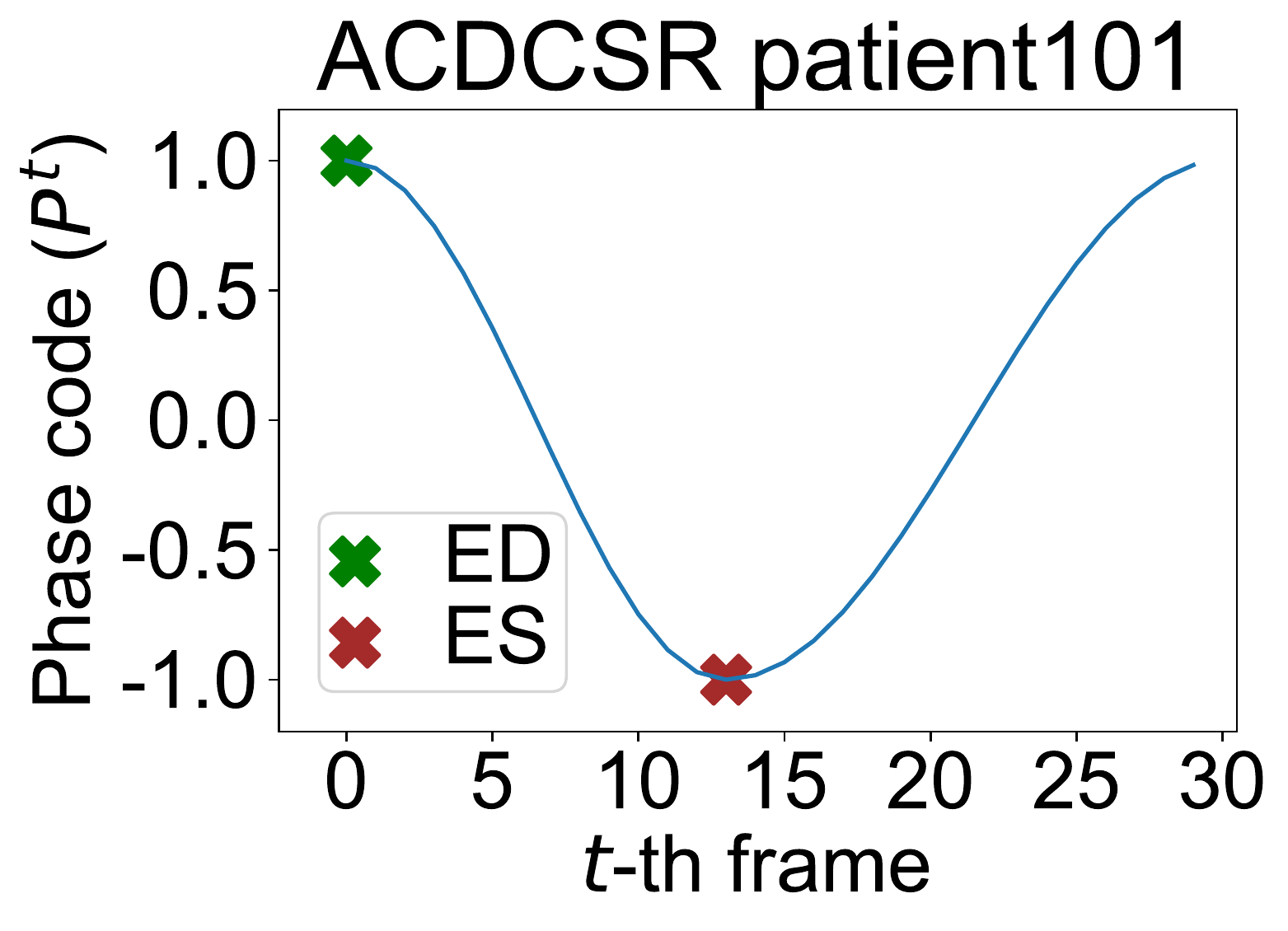}
            \end{subfigure}
            \begin{subfigure}{\linewidth}
                \includegraphics[width=\linewidth]{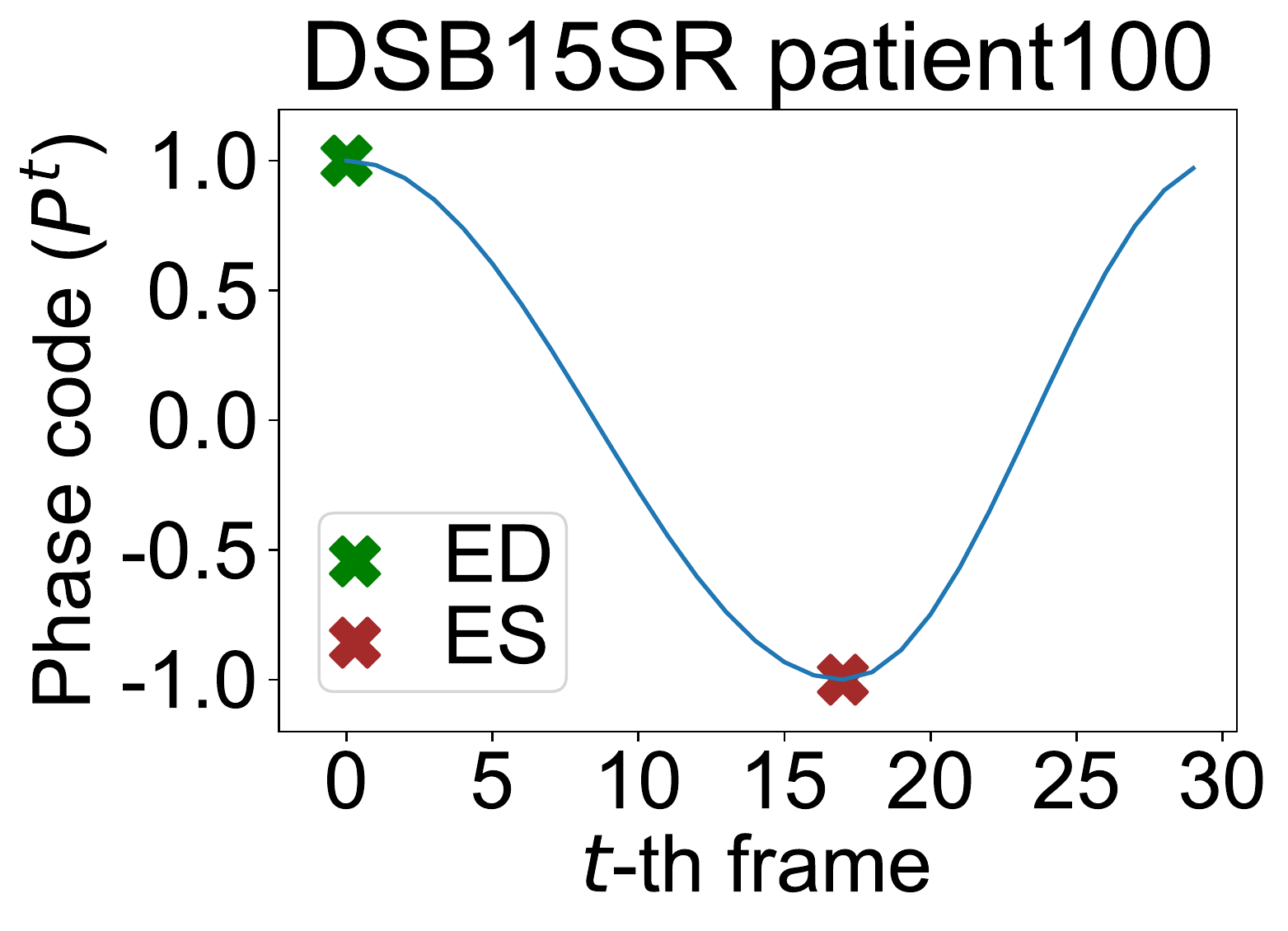}
            \end{subfigure}
            \vspace{-3pt}
            \caption{Phase code}
            \label{fig:phase code}
        \end{subfigure}
    \end{subfigure}
    \begin{subfigure}{0.31\linewidth}
        \centering
        \begin{subfigure}{\linewidth}
            \includegraphics[width=\linewidth,trim={7cm 1.3cm 6.8cm 1cm},clip]{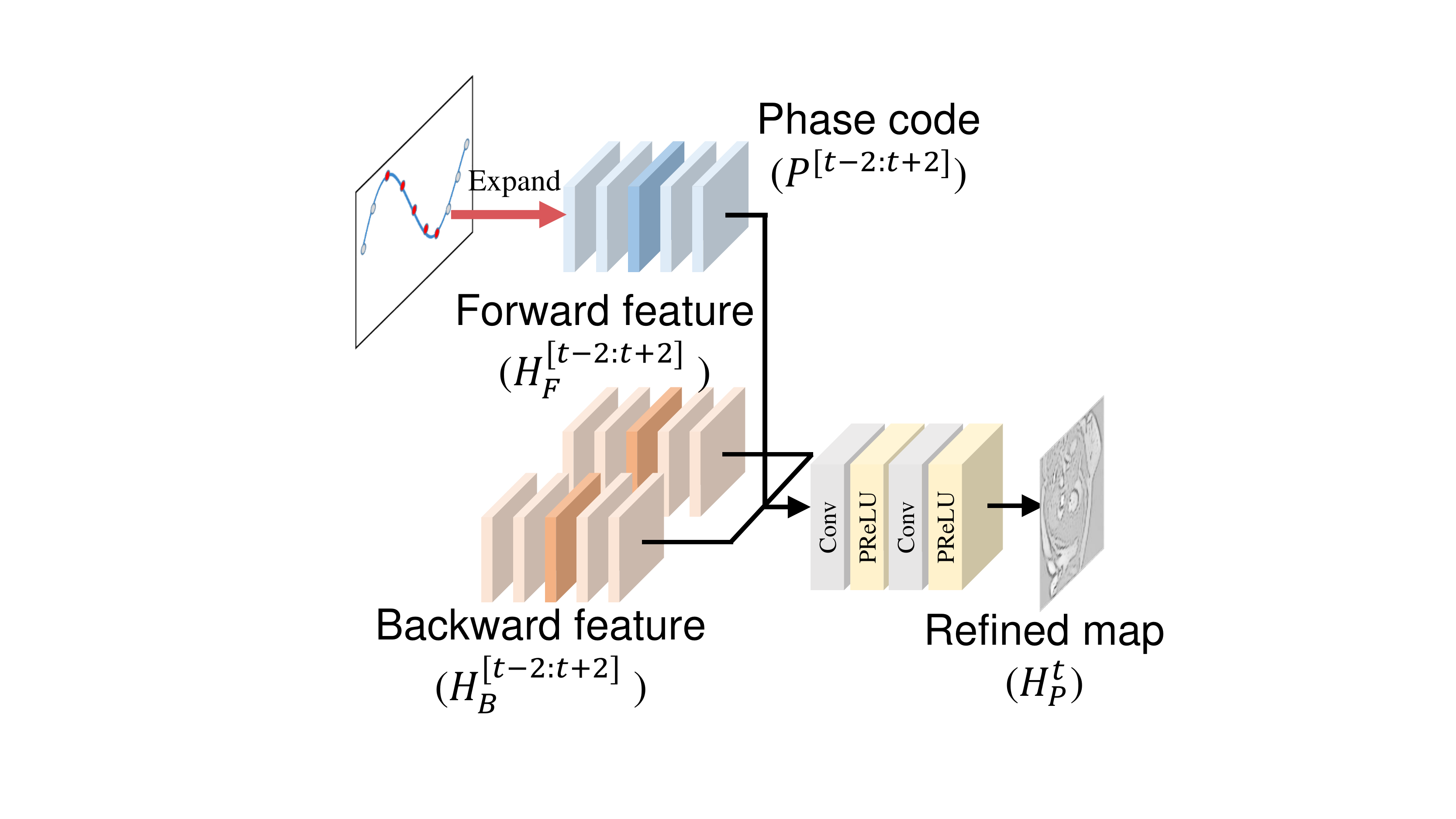}
            \caption{Phase fusion module}
            \label{fig:phase fusion module}
        \end{subfigure}
    \end{subfigure}
    \begin{subfigure}{0.49\linewidth}
        \includegraphics[width=\linewidth,trim={8.5cm 3.5cm 8.5cm 3.5cm},clip]{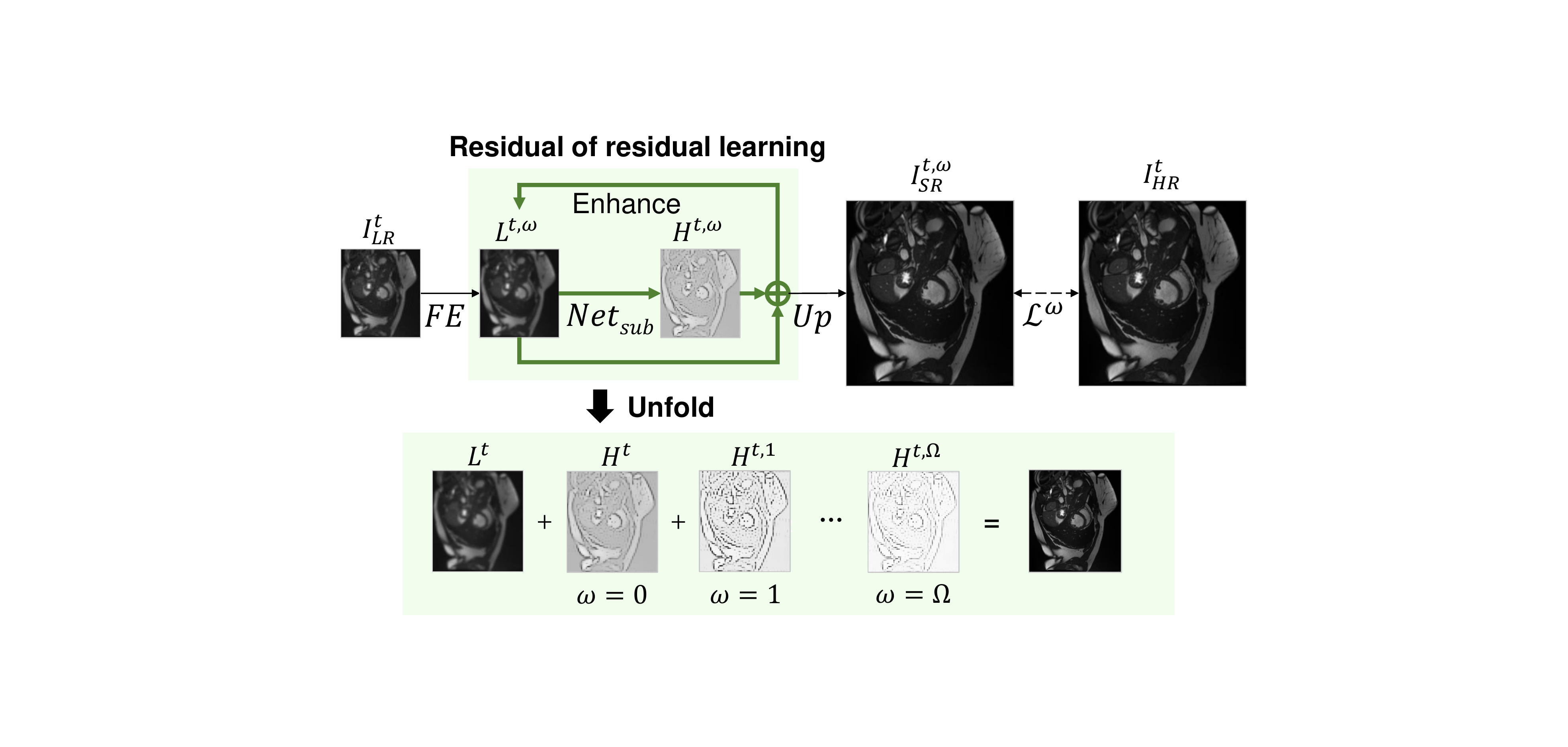}
        \caption{Residual of residual learning}
        \label{fig:residual of residual learning}
    \end{subfigure}
    \caption{\textbf{Proposed components.}
    (a) \emph{Phase code} formulated as the periodic function contains domain knowledge (\ie, cardiac phase). 
    (b) \emph{Phase fusion module} can realize the phase of the current sequence with the cardiac knowledge to thoroughly integrate the bidirectional features.
    (c) \emph{Residual of residual learning} aims at directing the model to reconstruct the results in a coarse-to-fine manner.
    }
\end{figure}

\subsection{Phase fusion module}
\label{sec:phase fusion module}
The cardiac cycle is a cyclic sequence of events when the heart beats, which consists of systole and diastole process.
Identification of the end-systole ($ES$) and the end-diastole ($ED$) in a cardiac cycle has been proved critical in several applications, such as the measurement of the ejection fraction and stroke volume~\cite{lalande2004left}, and disease diagnosis~\cite{xu2017volume}.
Hence, we embed the physical meaning of the input frames into our model with the informative phase code generated by projecting the cardiac cycle to the periodic Cosine function as depicted in Fig.~\ref{fig:phase code}.
Specifically, we map the process of the systole and the diastole to the half-period cosine separately:
\begin{equation}
    P^t = \begin{cases}
        Cos(\pi \times \frac{t-ED}{ES-ED}), & \text{if } \; \text{ED} < t \leq \text{ES}\\
        Cos(\pi \times (1+\frac{(t-ES)\%T}{T-(ES-ED)})), & \text{otherwise}
    \end{cases}
\end{equation}
where \% denotes modulo operation and $T$ is the frame number in a cardiac cycle.
The overview of the proposed phase fusion module is shown in Fig~\ref{fig:phase fusion module}.
The features from the bidirectional ConvLSTM with the corresponding phase code are concatenated and fed into the fusion module.
With the help of consecutive $2N+1$ phase codes, it can link the same-position frames from different periods (inter-period).
Besides, it can realize the heart is relaxing or contracting as the phase code is respectively increasing or decreasing (intra-period).

\subsection{Residual of residual learning}
\label{sec:residual of residual learning}
In the computer vision field, the iterative error-correcting mechanism plays an essential role in several topics, such as reinforcement learning~\cite{mnih2013playing}, scene reconstruction~\cite{montemerlo2002fastslam}, and human pose estimation~\cite{carreira2016human}.
Inspired by this mechanism, we propose the residual of residual learning composing the reconstruction process into multiple stages, as shown in Fig.~\ref{fig:residual of residual learning}.
At each stage, the sub-network ($Net_{sub}$) in our model estimates the high-frequency residual based on the current low-frequency feature, and then the input low-frequency feature is updated for the next refinement stage.
Let $L^{t, 0}$ be the initial feature from the feature extractor ($FE$) and $L^{t, \omega}$ denote the updated feature at the iteration $\omega$, the residual of residual learning for $\Omega$ stages can be described as the recursive format:
\begin{equation}
    L^{t, \omega} = 
        \begin{cases}
            FE(I^t_{LR}), & \text{when }\omega = 0
            \\
            L^{t, \omega-1} \oplus Net_{sub}(L^{t, \omega-1}), & \text{if } 0 < \omega \leq \Omega
        \end{cases}
\end{equation}
Then, the network generates the super-resolution result $I^{t, \omega}_{SR}$ based on the current reconstructed feature $L^{t, \omega}$, which can be written as:
\begin{equation}
    I^{t, \omega}_{SR} = Up(L^{t, \omega} \oplus Net_{sub}(L^{t, \omega}))
\end{equation}
The model progressively restores the residual that has yet to be recovered in each refinement stage, which is so-called the residual of residual learning.
Compared to other one-step approaches~\cite{lim2017enhanced,jo2018deep,wang2019edvr,xue2019video,sajjadi2018frame}, the proposed mechanism tries to break down the ill-posed problem into several easier sub-problems in the manner of divide-and-conquer.
Most notably, it can dynamically adjust the iteration number depending on the problem difficulty without any additional parameters.

\subsection{Loss function}
\label{sec:loss function}
In this section, we elaborate on the mathematical formulation of our cost function.
At each refinement stage $\omega$, the super-resolved frames \{$I^{t, \omega}_{SR}$\} are supervised by the ground-truth HR video \{$I^t_{HR}$\}, which can be formulated as $\mathcal{L}^\omega = \frac{1}{\tilde{T}}\sum_{t=1}^{\tilde{T}}\ \parallel I^{t, \omega}_{SR} - I^t_{HR} \parallel_1$, where $\tilde{T}$ indicates the length of the video sequence fed into the network.
We choose the L1 loss as the cost function since the previous works have demonstrated that the L1 loss provides better convergence compared to the widely used L2 loss~\cite{zhao2015loss,lim2017enhanced}.
Besides, we apply the deep supervision technique as described in Sec.~\ref{sec:overall architecture} by adding two auxiliary losses $\mathcal{L}_F^\omega = \frac{1}{\tilde{T}}\sum_{t=1}^{\tilde{T}}\ \parallel I^{t, \omega}_{SR, F} - I^t_{HR} \parallel_1$ and $\mathcal{L}_B^\omega = \frac{1}{\tilde{T}}\sum_{t=1}^{\tilde{T}}\ \parallel I^{t, \omega}_{SR, B} - I^t_{HR} \parallel_1$.
Hence, the total loss function can be summarized as $\mathcal{L} = \sum_{\omega=0}^{\Omega} (\mathcal{L}^\omega + \mathcal{L}_F^\omega + \mathcal{L}_B^\omega)$,
where $\Omega$ denoted as the total number of refinement stages.

\begin{table*}[t]
    \caption{\textbf{Quantitative results.}
    The red and blue indicate the best and the second-best performance, respectively.
    We adopt CardiacPSNR/CardiacSSIM to fairly assess the reconstruction quality of the heart region.
    It is worth noting that the large-scale DSB15SR dataset is entirely for external evaluation.
    }
    \label{tab:cardiac}
    \centering
    \resizebox{\textwidth}{!}{
        \begin{tabular}{ccccccccccc}
            \toprule
            \multirow{2}[5]{*}{Dataset} & \multirow{2}[5]{*}{Scale} &
            \multicolumn{2}{c}{SISR} & \multicolumn{6}{c}{VSR} \\
            \cmidrule(lr){3-4} \cmidrule(lr){5-10}
             & & Bicubic & EDSR\cite{lim2017enhanced} & DUF\cite{jo2018deep} & EDVR\cite{wang2019edvr} & RBPN\cite{haris2019recurrent} & TOFlow\cite{xue2019video} & FRVSR\cite{sajjadi2018frame} & \makecell[c]{Model \\ (Ours)} \\
            \midrule
            \multirow{3}{*}{ACDCSR} & $\times 2$ & 33.0927 / 0.9362 & 37.3022 / 0.9681 & 37.4008 / 0.9688 & - / - & \textcolor{red}{37.5017} / \textcolor{blue}{0.9694} & 36.6510 / 0.9641 & - / - & \textcolor{blue}{37.5003} / \textcolor{red}{0.9696} \\
            & $\times 3$ & 29.0724 / 0.8472 & 32.8177 / 0.9201 & 32.7942 / 0.9203 & - / - & \textcolor{blue}{32.9099} / \textcolor{blue}{0.9225} & 32.4535 / 0.9136 & - / - & \textcolor{red}{32.9342} / \textcolor{red}{0.9231} \\
            & $\times 4$ & 26.9961 / 0.7611 & 30.2536 / 0.8631 & 30.2420 / 0.8621 & 30.2817 / \textcolor{blue}{0.8655} & \textcolor{blue}{30.3294} / 0.8653 & 30.0087 / 0.8538 & 30.1693 / 0.8592 & \textcolor{red}{30.4060} / \textcolor{red}{0.8668} \\
            \midrule
            \multirow{3}{*}{DSB15SR} & $\times 2$ & 34.1661 / 0.9597 & 40.1723 / 0.9815 & 40.3548 / \textcolor{blue}{0.9822} & - / - & \textcolor{blue}{40.3792} / \textcolor{red}{0.9824} & 39.5042 / 0.9794 & - / - & \textcolor{red}{40.4635} / 0.9821 \\
            & $\times 3$ & 29.1175 / 0.8854 & 33.9893 / 0.9424 & 33.9736 / 0.9428 & - / - & \textcolor{blue}{34.1320} / \textcolor{blue}{0.9445} & 33.6656 / 0.9386 & - / - & \textcolor{red}{34.2169} / \textcolor{red}{0.9451} \\
            & $\times 4$ & 26.5157 / 0.8065 & 30.6354 / 0.8907 & 30.7411 / 0.8918 & \textcolor{blue}{30.8564} / \textcolor{blue}{0.8949} & 30.7985 / 0.8933 & 30.3153 / 0.8836 & 30.5800 / 0.8889 & \textcolor{red}{30.9104} / \textcolor{red}{0.8956}\\
            \bottomrule
      \end{tabular}
    }
\end{table*}

\section{Experiment}
\subsection{Experimental settings}

\subsubsection{Data preparation}
\label{sec:data preparation}
To our best knowledge, there is no publicly available CMR dataset for the VSR problem.
Hence, we create two datasets named ACDCSR and DSB15SR based on the public MRI datasets.
One is the Automated Cardiac Diagnosis Challenge dataset~\cite{bernard2018deep}, which contains four dimension MRI scans of a total of 150 patients.
The other is the large-scale Second Annual Data Science Bowl Challenge dataset~\cite{datasciencebowl2015} composed of 2D cine MRI videos that contain 30 images across the cardiac cycle per sequence.
We use its testing dataset comprising 440 patients as the external assessment to verify the robustness and generalization of the algorithms.
To more accurately mimic the acquisition of LR MRI scans~\cite{chen2018brain,zhao2018self}, we project the HR MRI videos to the frequency domain by Fourier transform and filter the high-frequency information.
After that, we apply the inverse Fourier transform to project the videos back to the spatial domain and further downsample by bicubic interpolation with the scale factor 2, 3, and 4.

\subsubsection{Evaluation metrics}
\label{sec:evalutaion metrics}
PSNR and SSIM criteria have been widely used in previous studies to evaluate the SR algorithms.
However, the considerable disparity of the proportion of the cardiac region to the background region in MRI images makes the results heavily biased towards the insignificant background region.
Therefore, we introduce CardiacPSNR and CardiacSSIM to assess the performance more impartially and objectively.
Specifically, we employ a heart ROI detection method similar to~\cite{tautz2011automatic} to crop the cardiac region and calculate PSNR and SSIM in this region.
This can reduce the influence of the background region and more accurately reflect the reconstruction quality of the heart region.

\subsubsection{Training details}
\label{sec:training details}
For training, we randomly crop the LR clips of $\tilde{T} = 7$ consecutive frames of size $32\times32$ with the corresponding HR clips.
We experimentally choose $n = 6$ and $\Omega = 2$ as detailed in Sec.~\ref{sec:ablation study}, while $N = 2$ in the phase fusion module.
We use the Adam optimizer~\cite{kingma2014adam} with learning rate $10^{-4}$ and set the batch size to $16$.
For other baselines, we basically follow their original settings except the necessary modifications to train them from the scratch.

\begin{figure}[t]
    \centering
    \begin{subfigure}{0.4455\linewidth}
        \includegraphics[width=\linewidth]{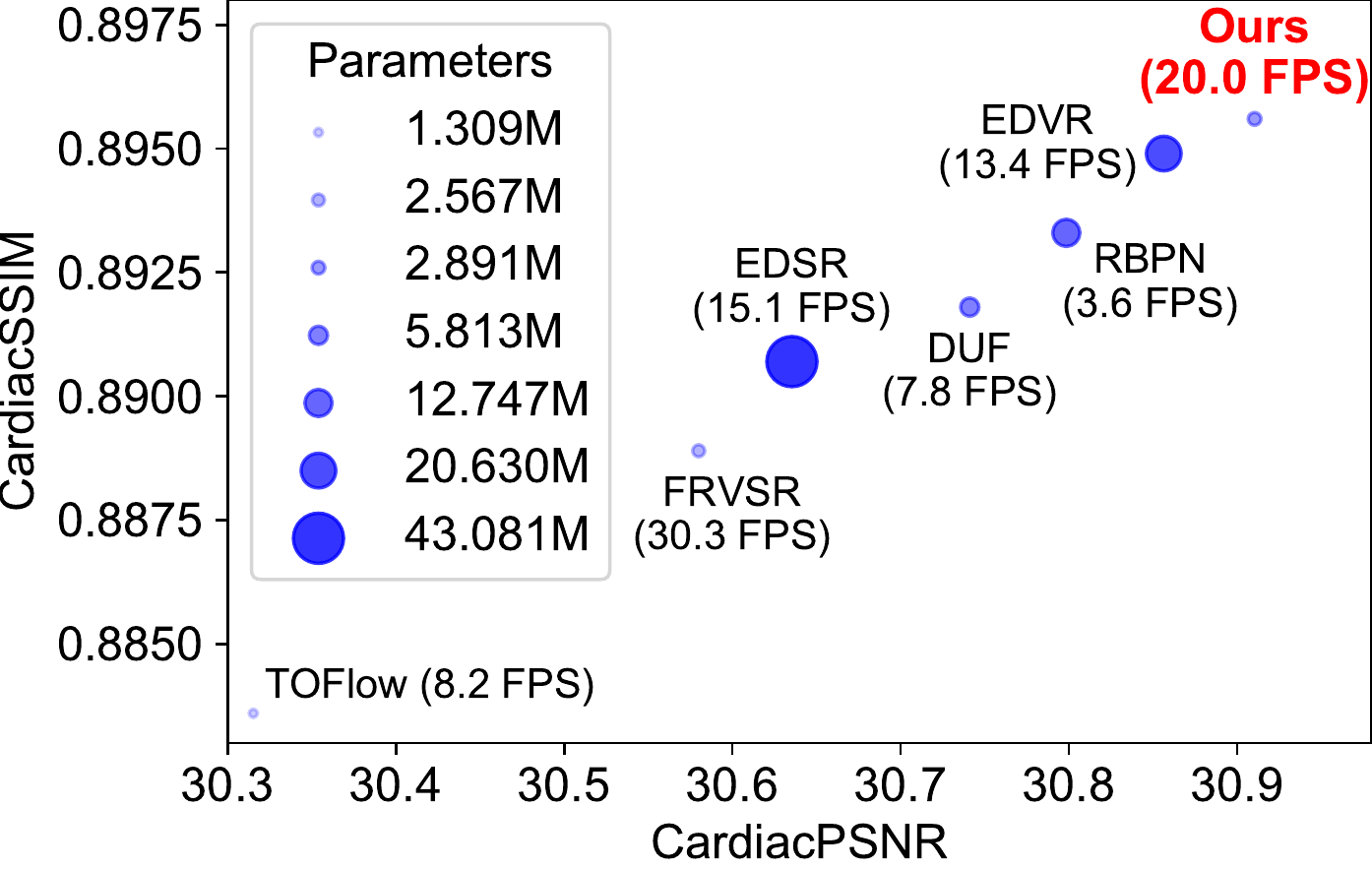}
        \caption{
        Efficiency vs performance on DSB15SR dataset for scale $\times 4$.
        (FPS: processed frames per second)}
        \label{fig:parameters}
    \end{subfigure}
    \quad
    \begin{subfigure}{0.52\linewidth}
        \begin{subfigure}{\linewidth}
            \begin{subfigure}{0.45\linewidth}
                \includegraphics[width=\linewidth]{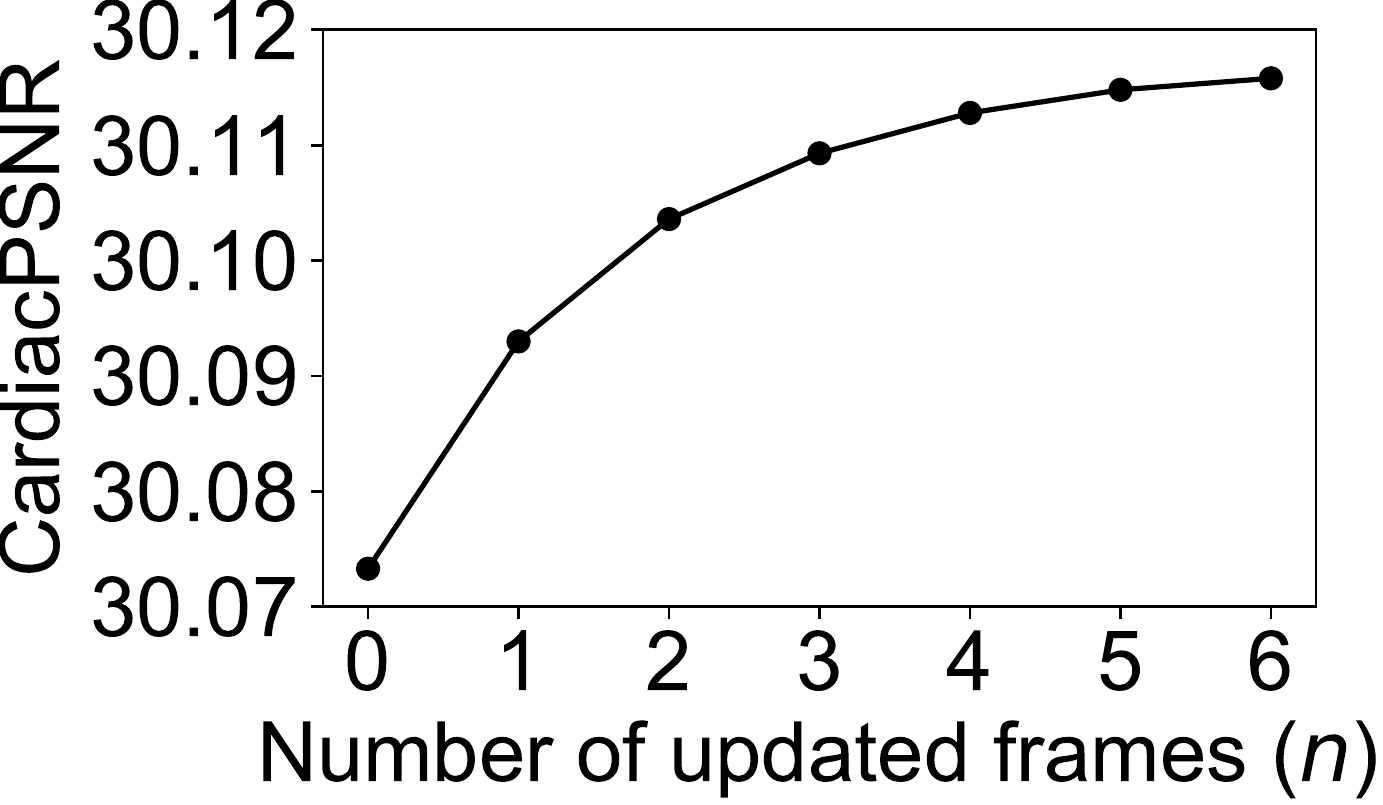}
            \end{subfigure}
            \quad
            \begin{subfigure}{0.45\linewidth}
                \includegraphics[width=\linewidth]{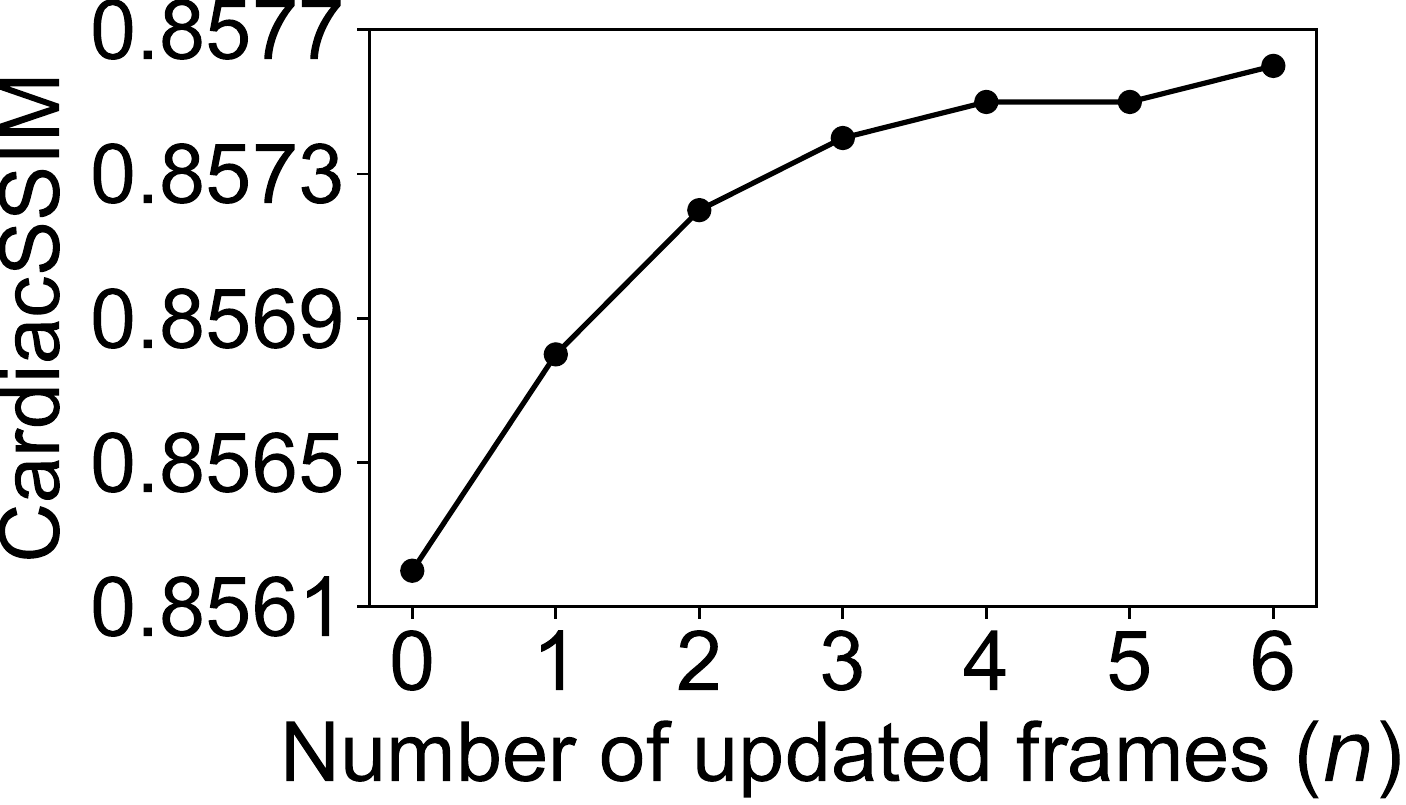}
            \end{subfigure}
            \caption{Analysis of the update frame number $n$.}
            \vspace{2mm}
            \label{fig:updated_frames}
        \end{subfigure}
        \\
        \begin{subfigure}{\linewidth}
            \begin{subfigure}{0.45\linewidth}
                \includegraphics[width=\linewidth]{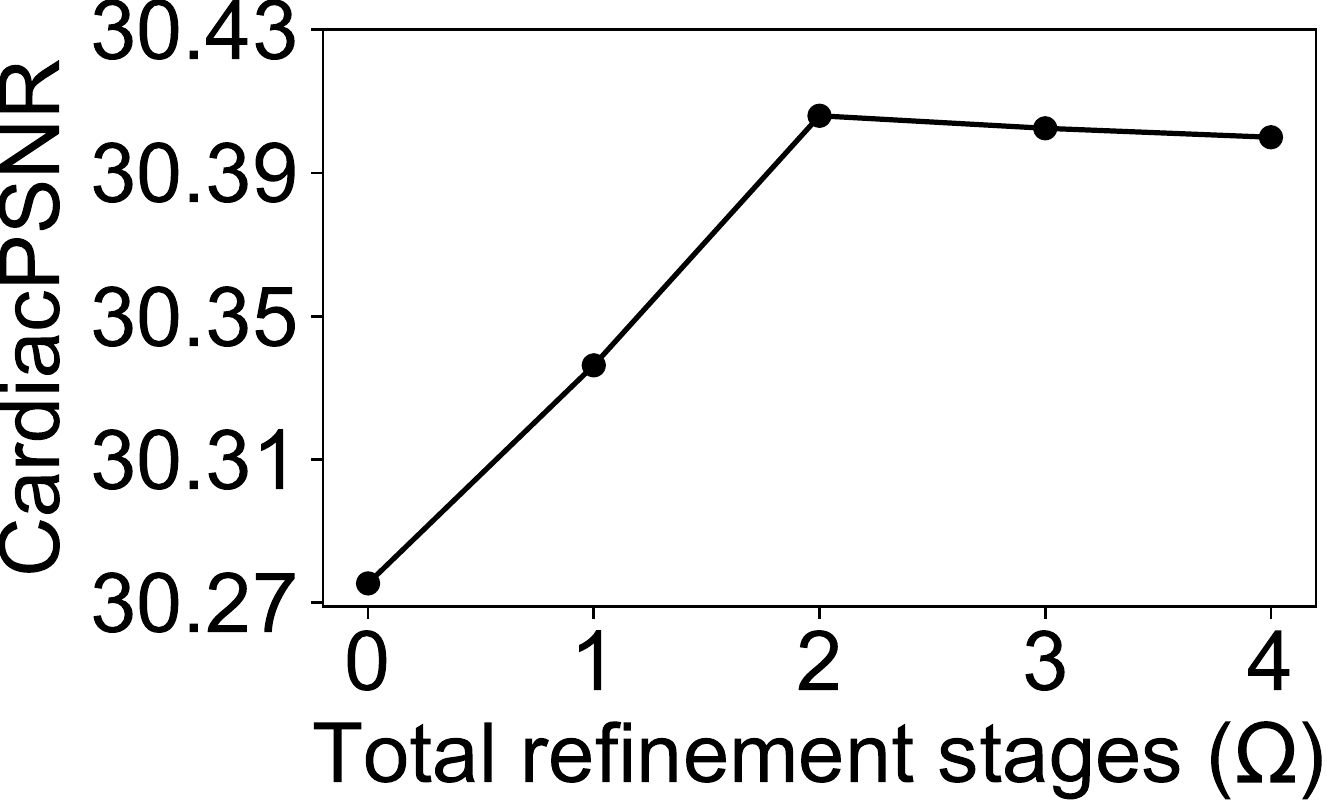}
            \end{subfigure}
            \quad
            \begin{subfigure}{0.45\linewidth}
                \includegraphics[width=\linewidth]{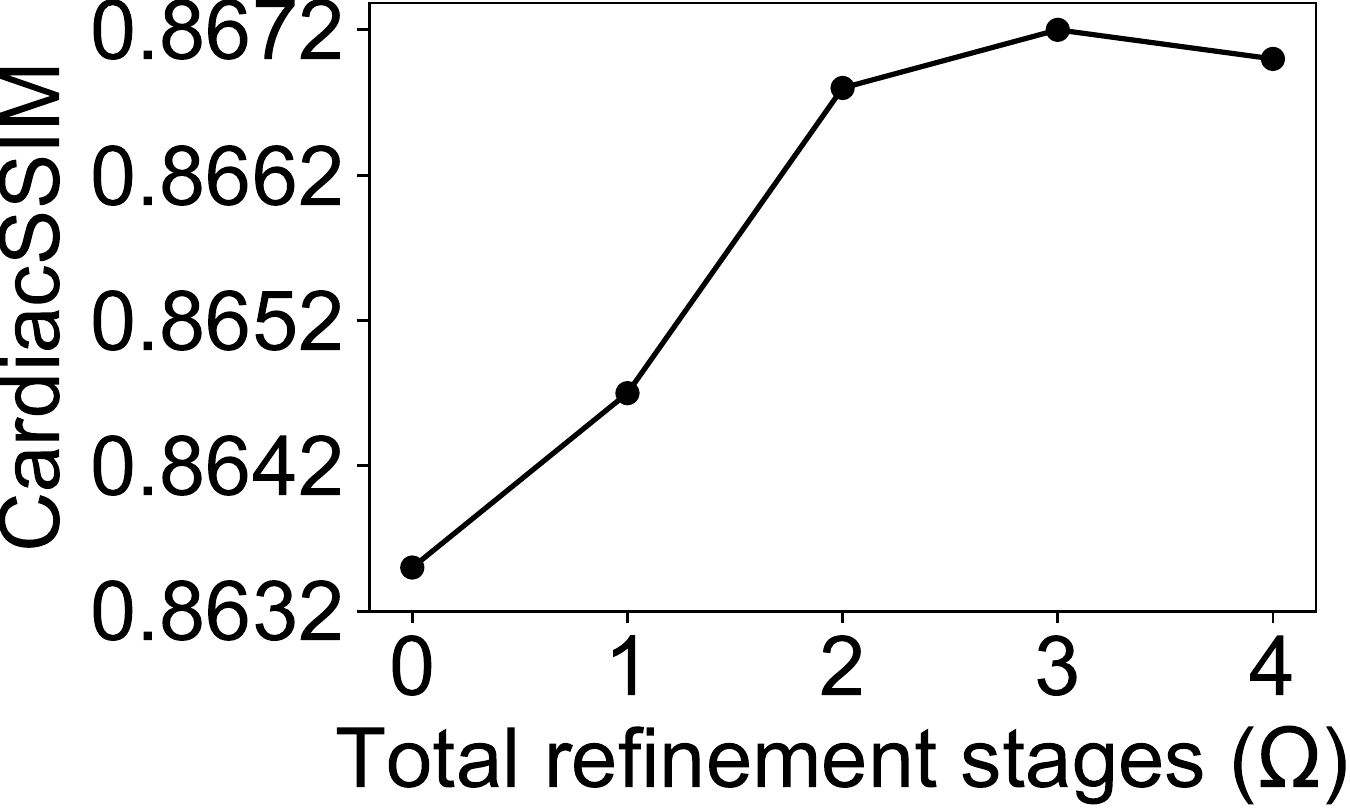}
            \end{subfigure}
            \caption{Analysis of total refinement stages $\Omega$.}
            \label{fig:refinements}
        \end{subfigure}
    \end{subfigure}
    \caption{\textbf{Experimental analysis.}
    (a) Our network outperforms other baselines with fewer parameters and higher FPS.
    (b) The performance is progressively enhanced as $n$ increases, which indicates that the prior sequence can provide useful information.
    (c) The performance can be improved with $\Omega$ increasing.
    }
\end{figure}

\begin{table*}[b]
    \caption{
        \textbf{Ablation study.}
        \emph{Memory}: the memory cells in the ConvLSTM~\cite{xingjian2015convolutional} are activated; \emph{Updated memory}: the memory cells are updated by feeding $n$ consecutive frames; \emph{Bidirection}: bidirectional ConvLSTM is adopted; \emph{Phase fusion module} and \emph{Residual of residual learning}: the proposed components are adopted.
    }
    \label{tab:ablation study}
    \centering
    \resizebox{\textwidth}{!}{
        \begin{tabular}{cccccc}
            \toprule
                Memory & \makecell[c]{Updated memory \\ ($n$ = 6)} & Bidirection & \makecell[c]{Phase fusion module} & \makecell[c]{Residual of residual learning \\ ($\Omega = 2$)} & CardiacPSNR/CardiacSSIM\\
                \midrule
                & & & & & 29.7580 / 0.8458\\
                \checkmark & & & & & 30.0733 / 0.8562\\
                \checkmark & \checkmark & & & & 30.1790 / 0.8596\\
                \checkmark & \checkmark & \checkmark & & & 30.2380 / 0.8623\\
                \checkmark & \checkmark & \checkmark & \checkmark & & 30.2754 / 0.8635\\
                \checkmark & \checkmark & \checkmark & \checkmark & \checkmark & \textbf{\textcolor{red}{30.4060}} / \textbf{\textcolor{red}{0.8668}}\\
            \bottomrule
      \end{tabular}
    }
\end{table*}

\subsection{Experimental results}
\begin{figure*}[t]
    \captionsetup[subfigure]{labelformat=empty}
    \centering
    \begin{subfigure}{0.9\linewidth}
        \centering
        \begin{subfigure}{0.04\linewidth}
            \vspace{2.5mm}
            \caption{$\times 3$}
        \end{subfigure}
        \quad
        \begin{subfigure}{0.175\linewidth}
            \includegraphics[width=\linewidth]{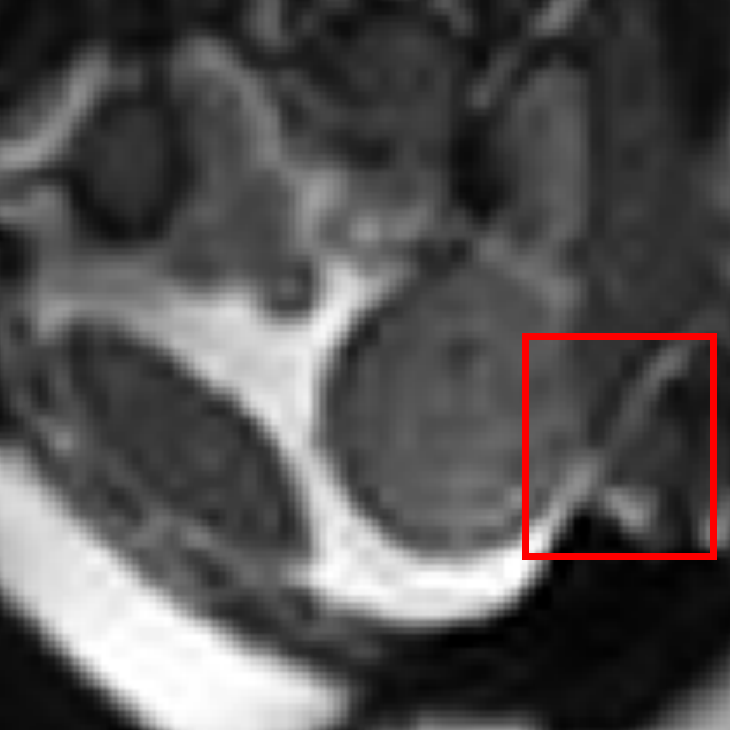}
        \end{subfigure}
        \begin{subfigure}{0.175\linewidth}
            \includegraphics[width=\linewidth]{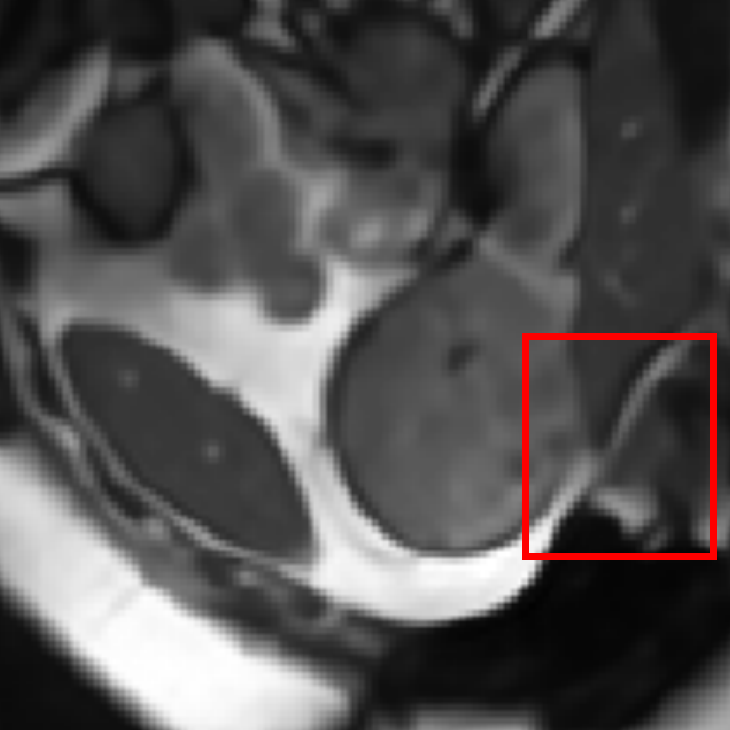}
        \end{subfigure}
        \begin{subfigure}{0.175\linewidth}
            \includegraphics[width=\linewidth]{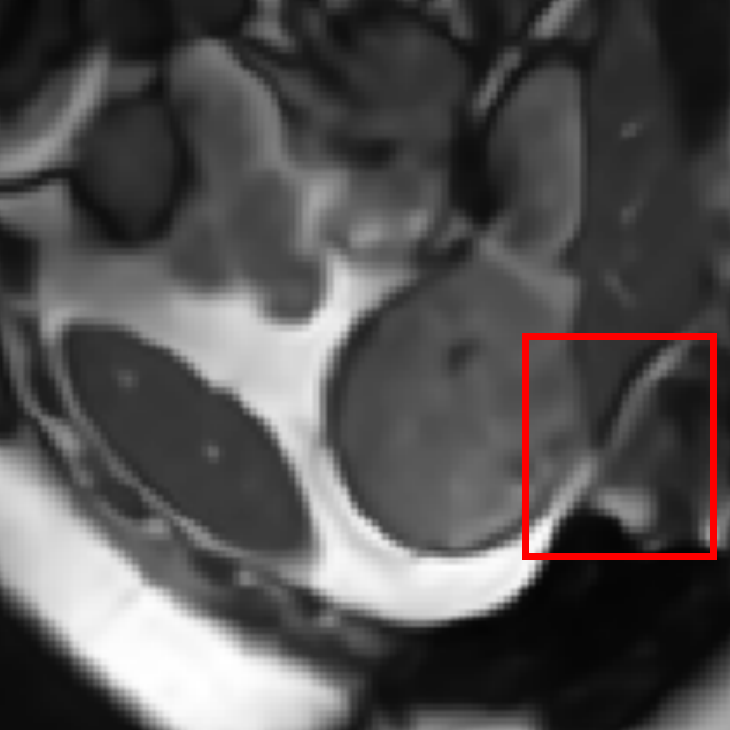}
        \end{subfigure}
        \begin{subfigure}{0.175\linewidth}
            \includegraphics[width=\linewidth]{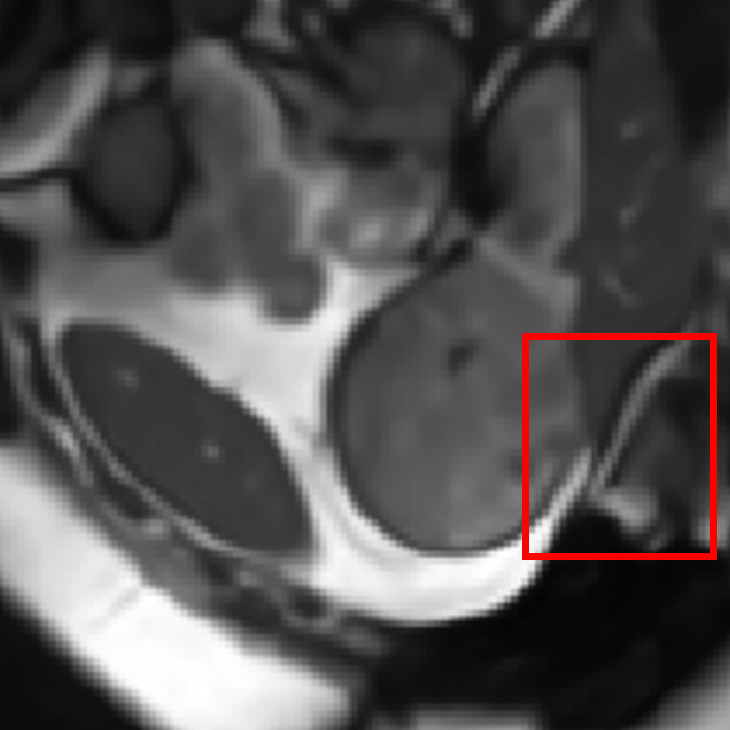}
        \end{subfigure}
        \begin{subfigure}{0.175\linewidth}
            \includegraphics[width=\linewidth]{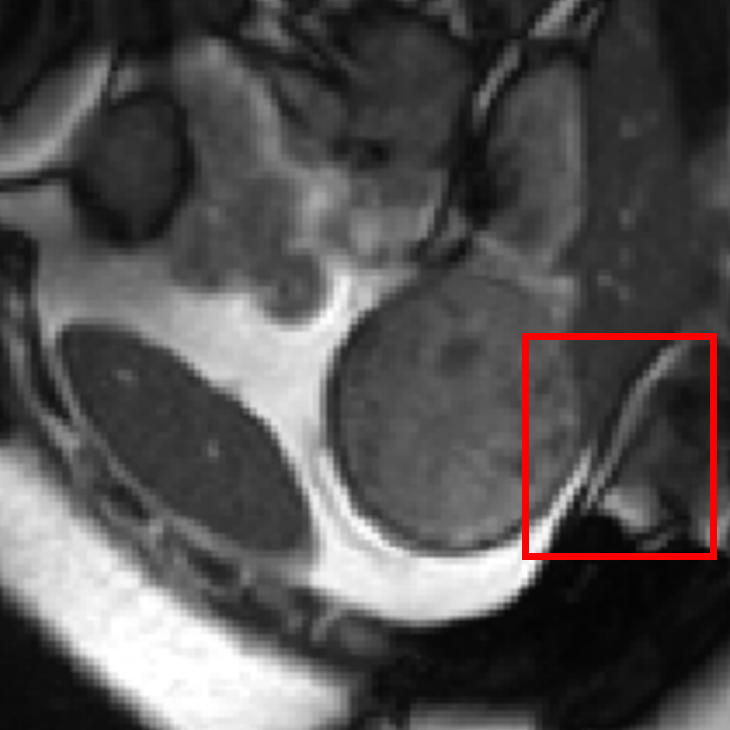}
        \end{subfigure}
    \end{subfigure}
    \begin{subfigure}{0.9\linewidth}
        \centering
        \begin{subfigure}{0.04\linewidth}
            \caption{$\times 4$}
            \vspace{4mm}
        \end{subfigure}
        \quad
        \begin{subfigure}{0.175\linewidth}
            \includegraphics[width=\linewidth]{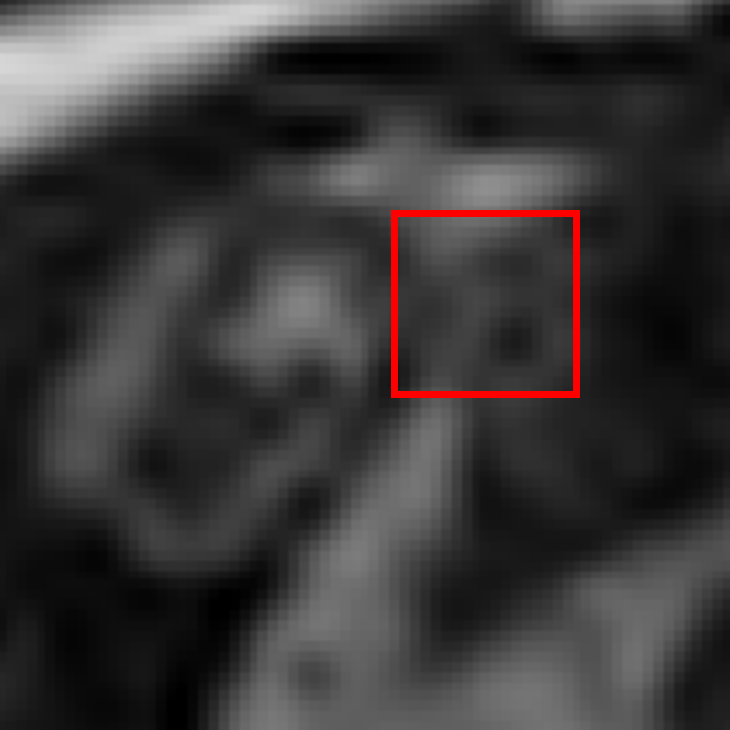}
            \caption{Bicubic}
        \end{subfigure}
        \begin{subfigure}{0.175\linewidth}
            \includegraphics[width=\linewidth]{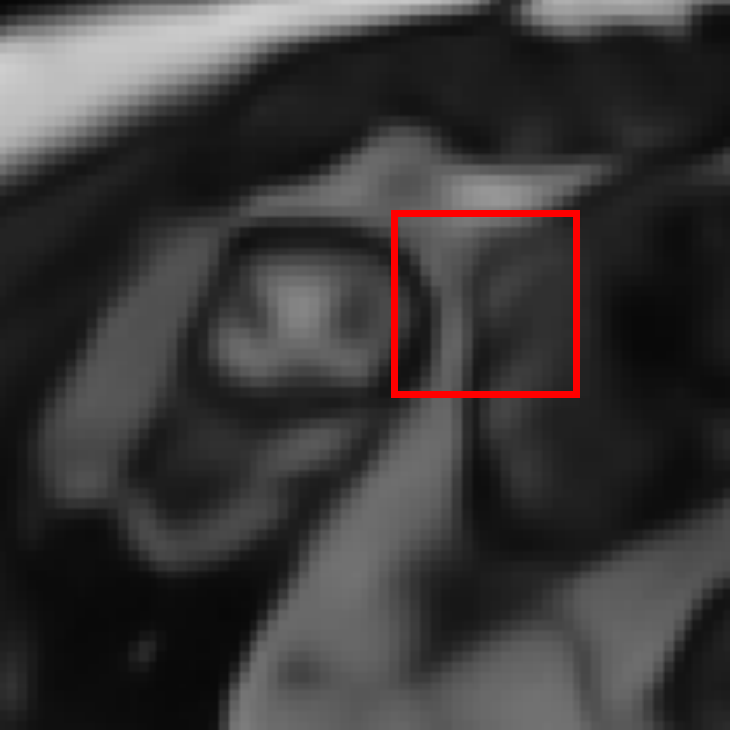}
            \caption{EDSR~\cite{lim2017enhanced}}
        \end{subfigure}
        \begin{subfigure}{0.175\linewidth}
            \includegraphics[width=\linewidth]{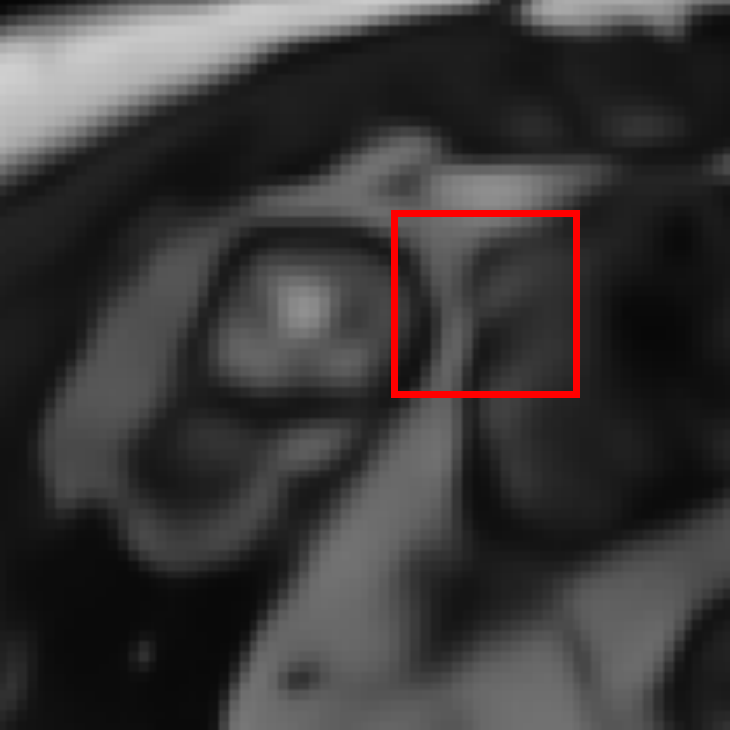}
            \caption{RBPN~\cite{haris2019recurrent}}
        \end{subfigure}
        \begin{subfigure}{0.175\linewidth}
            \includegraphics[width=\linewidth]{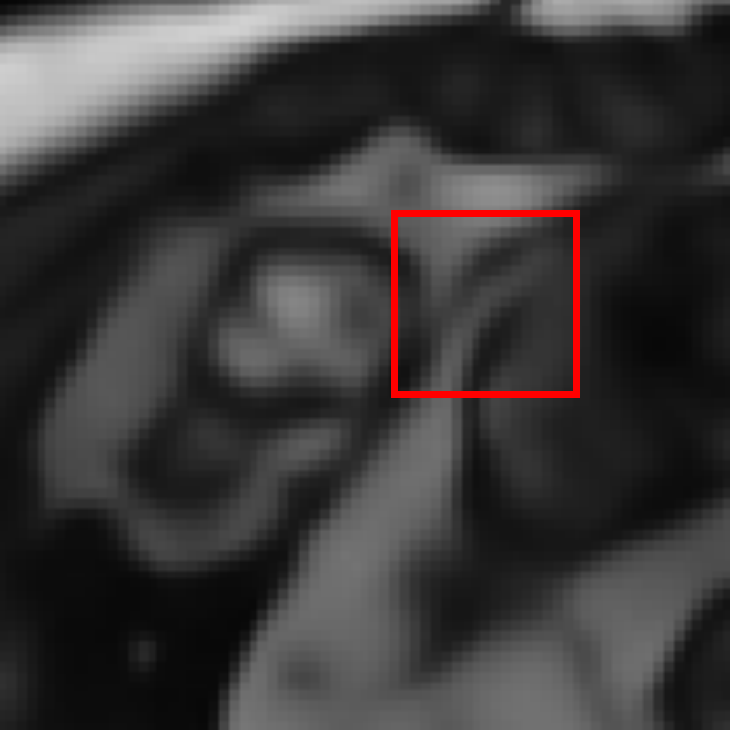}
            \caption{Ours}
        \end{subfigure}
        \begin{subfigure}{0.175\linewidth}
            \includegraphics[width=\linewidth]{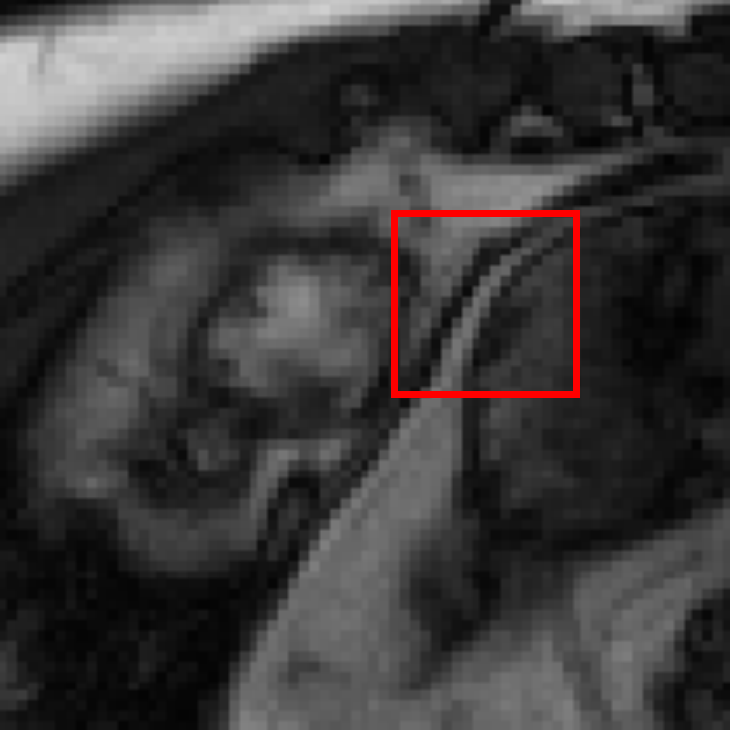}
            \caption{HR}
        \end{subfigure}
    \end{subfigure}
    \caption{
        \textbf{Qualitative results.}
        Zoom in to see better visualization.
    }
    \label{fig:visualizations}
\end{figure*}

To confirm the superiority of the proposed approach, we compare our network with multiple state-of-the-art methods, namely EDSR~\cite{lim2017enhanced}, DUF~\cite{jo2018deep}, EDVR~\cite{wang2019edvr}, RBPN~\cite{haris2019recurrent}, TOFlow~\cite{xue2019video}, and FRVSR~\cite{sajjadi2018frame}.
We present the quantitative and qualitative results in Tab.~\ref{tab:cardiac} and Fig.~\ref{fig:visualizations} respectively.
Our approach outperforms almost all the existing methods by a huge margin in all scales in terms of CardiacPSNR and CardiacSSIM.
In addition, our method can yield more clear and photo-realistic SR results which subjectively closer to the ground truths.
Moreover, the results on the external DSB15SR dataset are sufficiently convincing to validate the generalization of the proposed approach.
On the other hand, the comparison with regard to the model parameters, FPS, and the image quality in the cardiac region plotted in Fig.~\ref{fig:parameters} demonstrates that our method strikes the best balance between efficiency and reconstruction performance.

\subsection{Ablation study}
\label{sec:ablation study}
We adopt the unidirectional ConvLSTM as the simplest baseline.
As shown in the Tab.~\ref{tab:ablation study}, the temporal information is important since the model performance is worse when the memory cells in ConvLSTM are disabled.
As the cardiac MRI video is cyclic, we can refresh the memory by feeding $n$ successive frames.
Accordingly, we analyze the relation between $n$ and model performance.
The result in Fig~\ref{fig:updated_frames} turns out that the network significantly improves as the updated frame number increases.
Moreover, the forward and backward information is shown to be useful and complementary for recovering the lost details.
In Sec.~\ref{sec:phase fusion module}, we exploit the knowledge of the cardiac phase to better fuse the bidirectional information.
The result in Tab.~\ref{tab:ablation study} reveals that the phase fusion module can leverage the bidirectional temporal features more effectively.
Besides, we explore the influence of the total number of refinement stages $\Omega$ in the residual of residual learning.
It can be observed from Fig.~\ref{fig:refinements} that the reconstruction performance is improved as the total refinement stages continue to increase.
The possible reason for the saturation or degradation of the overall performance when $\Omega$ equals to 3 or 4 is overfitting (violate the Occam’s razor).

\section{Conclusion}
In this work, we define the cyclic cardiac MRI video super-resolution problem which has not yet been completely solved to our best knowledge.
To tackle this issue, we bring the cardiac knowledge into our network and employ the residual of residual learning to train in the progressive refinement manner, which enables the model to generate sharper results with fewer model parameters.
In addition, we build large-scale datasets and introduce cardiac metrics for this problem.
Through extensive experiments, we demonstrate that our network outperforms the state-of-the-art baselines qualitatively and quantitatively.
Most notably, we carry out the external evaluation, which indicates our model exhibits good generalization behavior.
We believe our approach can be seamlessly applied to other modalities such as computed tomography angiography and echocardiography.
\section{Acknowledgment}
This work was supported in part by the Ministry of Science and Technology, Taiwan, under Grant MOST 109-2634-F-002-032 and Microsoft Research Asia. We are grateful to the NVIDIA grants and the DGX-1 AI Supercomputer and the National Center for High-performance Computing. We thank Dr. Chih-Kuo Lee, National Taiwan University Hospital, for the early discussions.

\bibliographystyle{splncs04}
\bibliography{paper}

\end{document}